\documentclass[12pt]{article}
\begin{document}
\title
{Thermodynamics of a charged particle in a noncommutative plane in a background magnetic field}
\author{
{\bf {\normalsize Aslam Halder}$^{a}
$\thanks{aslamhalder.phy@gmail.com}},
{\bf {\normalsize Sunandan Gangopadhyay}
$^{b,c}$\thanks{sunandan.gangopadhyay@gmail.com, sunandan@iiserkol.ac.in, sunandan@associates.iucaa.in}},\\[0.2cm]
$^{a}$ {\normalsize Kolorah H.A.W Institution, Kolorah, Howrah-711411, India}\\[0.2cm]
$^{b}$ {\normalsize Department of Physical Sciences,}\\
{\normalsize Indian Institute of Science Education and Research Kolkata} \\
{\normalsize Mohanpur 741246, Nadia, West Bengal, India}\\[0.2cm]
$^{c}${\normalsize Visiting Associate in Inter University Centre for Astronomy $\&$ Astrophysics (IUCAA),}\\
{\normalsize Pune 411007, India}\\[0.3cm]
}
\date{}

\maketitle
\begin{abstract}
\noindent
Landau system in noncommutative space has been considered. 
To take into account the issue of gauge invariance in noncommutative space, we incorporate the Seiberg-Witten map in our analysis. Generalised Bopp-shift transformation is then used to map the noncommutative system to its commutative equivalent system. In particular we have computed the partition function of the system and from this we obtained the susceptibility of the Landau system and found that the result gets modified by the spatial noncommutative parameter $\theta$. We also investigate the de Hass--van Alphen effect in noncommutative space and observe that the oscillation of the magnetization and the susceptibility gets noncommutative corrections. Interestingly, the susceptibility in the noncommutative scenario is non-zero in the range of the magnetic field greater than the threshold value which is in contrast to its commutative counterpart. The results obtained are valid upto all orders in the noncommutative parameter $\theta$.   

\end{abstract}

\maketitle

\vskip 1cm
\section{Introduction}
It is well known that the quantum dynamics of a charged particle moving in a background homogeneous magnetic field, commonly known as the Landau problem \cite{landau} is directly linked with many quantum mechanical problems, namely, the integral and fractional quantum Hall effect \cite{ref_1, laugh_1}, Aharonov-Bohm effect \cite{aharonov}, 
anyons excitations in a rotating Bose-Einstein condensate \cite{ref_4, ref_5},  to name a few. 
The Landau problem also drew considerable attention in the context of the investigation of the thermodynamics of an electron moving in a constant external magnetic field. This led to the study of its magnetic properties and thereafter became well known as Landau diamagnetism \cite{landau}, \cite{feldman}.

On the other hand physics of a system living in a noncommutative (NC) space has drawn a lot of interest in recent years \cite{sw}-\cite{szabo}. 
Interestingly, the NC algebra on the two dimensional plane has a direct link with the Landau problem. It can be said that the Landau problem is perhaps the simplest example of space quantization. By explicit investigation one can find that the projection of coordinates to the lowest Landau level results in a NC algebra between the position coordinates of a particle in a two dimesional plane. This algebra is found to have a direct connection with the Moyal $\star$-product. NC geometry also arises in string theory with $D$-branes in background Neveu-Schwarz fields \cite{acny}-\cite{bcsgscholtz}. It has been shown that the $D$-brane world volume becomes a NC space and a low energy effective field theory can be arrived at in the point particle limit where the string length goes to zero. This leads to a noncommutative quantum field theory (NCQFT) \cite{sw}-\cite{szabo}, \cite{chams}-\cite{sghazra} where the NC coordinate algebra leads to an uncertainty in the spacetime geometry and the notion of a spacetime point gets replaced by a Planck cell of dimension given by the Planck area. 
Such NC geometrical structures are also known to arise in various theories of quantum gravity \cite{suss}-\cite{mof}. 

The low energy limit of NCQFT leads to noncommutative quantum mechanics (NCQM) and such theories have been studied extensively in the literature \cite{duval}-\cite{sgah}. All of these studies gives rise to NC corrections to the standard results. The importance of such studies lies in the fact that it leads to bounds on the NC parameter \cite{stern}-\cite{falomir}.


The primary aim of this paper is to study the thermodynamics  of an electron moving in a two dimensional NC plane in the presence of a background magnetic field perpendicular to the plane. This problem, popularly known as the Landau diamagnetism, has been studied much in the literature both in its commutative \cite{feldman} and NC formulations \cite{jellal}-\cite{vega}. Over the years this classic example has remained to be a continual area of intense research in the field of condensed matter physics. In the present work we revisit this problem once again. The point of departure of our analysis from the recent papers 
\cite{jellal, vega} is a careful attention being paid to the issue of gauge invariance. In the present work we have incorporated this issue by considering the Seiberg-Witten (SW) map. The SW map is a map which transforms the NC $U(1)_\star$ gauge invariant system to the usual commutative $U(1)$ gauge invariant system while preserving the physics. This map is obtained by demanding that the ordinary gauge potentials, which are connected by $U(1)$ gauge transformations, are mapped to NC potentials which are connected by the corresponding NC $U(1)_\star$ gauge transformations. The map that we use in this work is valid upto all orders in the NC parameter $\theta$. We have also used the generalised Bopp-shift transformation \cite{ah} to map the NC variables to their corresponding commuting counterpart. With this formalism in place,
we move on to compute the partition function of the Landau system by using the coherent state approach \cite{feldman}. We then calculate the magnetization and the susceptibility in the high temperature limit. We have found that the results get modified by the NC parameter $\theta$ and the corrections are captured upto all orders in $\theta$. We have then looked at the de Hass-van Alphen effect in this framework. 
Here we find that the oscillation of the magnetization and the susceptibility gets NC corrections. An important observation that we make here is that the susceptibility in the NC scenario is non-zero in the range of the magnetic field greater than the threshold value which is in contrast to its commutative counterpart \cite{jkb}.

         
This paper is organized as follows. In section 2, we present the preliminaries which involves introducing the problem of a charged particle moving in a two dimensional noncommutative plane in the presence of a constant background magnetic field. Here we discuss the necessary mapping relating the noncommutative and the commutative variables. The SW map is also introduced in this section. In section 3, the Landau diamagnetism in noncommutative space is investigated using a coherent state approach. In section 4, the de Hass--Van Alphen effect in noncommutative space is presented. Section 5 is devoted to concluding remarks.


\section{Preliminaries}
A nonrelativistic charged particle of mass $m$ moving in a two dimensional NC plane in the presence of a background magnetic field $\vec{B}(\theta)=\bar{B}(\theta)\hat{k}$ perpendicular to the plane is considered. Here $\theta$ is the spatial NC parameter and $\theta_{ij}$ is antisymmetric in the indices $i,j$, that is $\theta_{ij}=\theta \epsilon_{ij}$, where $\epsilon_{ij}=-\epsilon_{ji}, (\epsilon_{12}=1)$.
In general the NC algebra satisfied by the operators $\left(\hat{x}_i \, , \, \hat{p}_i\right)$ follow
\begin{eqnarray}
\label{e420}
\left[\hat{x}_{i},\hat{x}_{j}\right]=i\theta_{ij} ~ ; \quad \left[\hat{x}_{i},\hat{p}_{j}\right] = i\hbar\delta_{ij} ~ ;\quad \left[\hat{p}_{i},\hat{p}_{j}\right]=0~,~i,j=1,2.
\end{eqnarray}
Since the particle is moving in the $(x,y)$ plane, 
the motion along the $z$-direction is free and commutes with the $x$ and $y$ coordinates. The standard approach in the literature to deal with such problems is to form an equivalent commutative description of the NC theory by employing some transformation which relates the NC operators $\hat{x}_{i}$, $\hat{p}_{i}$ to ordinary commutative operators $x_{i}$, $p_{i}$ satisfying the usual Heisenberg algebra 
\begin{eqnarray}
\left[x_{i}, p_{j}\right]=i\hbar \delta_{ij}~; \quad \left[x_{i} \, , \, x_{j}\right]=0= \left[p_{i},p_{j}\right].
\label{cAlgebra}
\end{eqnarray}
The total Hamiltonian of the system is given by
\begin{eqnarray}
\label{e17a}
H=\hat{H}+H_{z}
\end{eqnarray}
where $\hat{H}$ is the Hamiltonian of the system for the motion of the charged particle in the NC $(x,y)$ plane  
\begin{eqnarray}
\label{e101}
\hat{H}=\frac{1}{2m}[(\vec{p}+\frac{e}{c}\hat{\vec{A}})\star(\vec{p}+\frac{e}{c}\hat{\vec{A}})]
\end{eqnarray}
and $H_{z}$ is the free particle Hamiltonian for the $z$-component of the particle 
\begin{eqnarray}
\label{e1011}
H_{z}=\frac{p_{z}^2}{2m}~.
\end{eqnarray}
The Schr\"{o}dinger equation of motion of the particle in the NC plane reads
\begin{eqnarray}
\label{e502}
i\frac{\partial \hat{\psi}(\hat{x},t)}{\partial t}=\hat{H}\star\hat{\psi}(\hat{x},t)=\hat{H}_{BS}(\theta)\hat{\psi}(\theta).
\end{eqnarray}
In the above equation first we have used star product \cite{mezin} and then the star product has been replaced by the Bopp-shift, defined in the usual way
\cite{mezin} 
\begin{eqnarray}
\label{e103}
(\hat{f}\star\hat{g})(x)=\hat{f}(x-\frac{\theta}{2}\epsilon^{ij}p_{j})\hat{g}(x).
\end{eqnarray} 
$\hat{H}_{BS}$ is the Bopp-shifted Hamiltonian which is in terms of commutative variables, however $\hat{\psi}$ appearing in eq.(\ref{e502}) is still noncommutative. In our subsequent discussion we shall use the generalised Bopp-shift transformations \cite{ah} which can be derived from the fact that the NC variables and commutative variables are related by the following set of equations
\begin{eqnarray}
\label{e74}
\hat{x}_{i}=a_{ij}x_{j}+b_{ij}p_{j}
\end{eqnarray}
\begin{eqnarray}
\label{e75}
\hat{p}_{i}=d_{ij}p_{j}
\end{eqnarray}
where $a$, $b$, and $d$ are $2\times 2$ transformation matrices. Imposing the NC Heisenberg algebra (\ref{e420}) and the usual Heisenberg algebra (\ref{cAlgebra}) on the above set of equations leads to the following set of conditions on the coefficients $a_{ij}$, $b_{ij}$, $d_{ij}$
\begin{eqnarray}
\label{e76}
ad^T=1
\end{eqnarray}
\begin{eqnarray}
\label{e77}
ab^T-ba^T=\frac{\theta}{\hbar}~.
\end{eqnarray}
We now assume $a_{ij}=\alpha\delta_{ij}$ and $d_{ij}=\beta\delta_{ij}$, where $\alpha$ and $\beta$ are two scaling constants . 
With these assumptions, eq.(\ref{e77}) give the solution for the matrix $b$ to be
\begin{eqnarray}
\label{e79}
b_{ij}=-\frac{1}{2\alpha\hbar}\theta_{ij}~.
\end{eqnarray}
Substituting this in eq.(\ref{e76}), we get
\begin{eqnarray}
\label{e8043}
\alpha\beta=1.
\end{eqnarray}
Hence the generalized Bopp-shift transformations read
\begin{eqnarray}
\label{e86}
\hat{x}_{i}  =   \alpha\left(x_{i}-\frac{1}{2\hbar\alpha^2}\theta_{ij}p_{j}\right)
\end{eqnarray}
\begin{eqnarray}
\label{e8zz} 
\hat{p}_{i}  = \frac{1}{\alpha}p_{i}~.
\end{eqnarray}
With the choice $\alpha=1$, the above set of transformations reduce to the well known Bopp-shift transformations \cite{mezin} 
\begin{eqnarray}
\label{e6}
\hat{x}_{i}  =   x_{i}-\frac{1}{2\hbar}\theta_{ij}p_{j}
\end{eqnarray} 
\begin{eqnarray}
\label{e6a} 
\hat{p}_{i}  =  p_{i}~.
\end{eqnarray}
To take into account $U(1)_{\star}$ gauge invariance, one needs to incorporate the SW map. It is a map from the NC space to commutative space which preserves the gauge invariance and the physics. To the lowest order in $\theta$, the SW maps for $\hat{\psi}$ and $\hat{A}_{k}$ read \cite{sw}, \cite{bcsgas}
\begin{eqnarray}
\label{e159}
\hat{\psi}=\psi-\frac{1}{2}\theta \epsilon^{ij}A_{i}\partial_{j}\psi
\end{eqnarray}
\begin{eqnarray}
\label{e163}
\hat{A_{k}}=A_{k}-\frac{1}{2}\theta \epsilon^{ij}A_{i}(\partial_{j}A_{k}+F_{jk}).
\end{eqnarray}
Before making further analysis, we first choose a gauge for the vector potentials $\hat{A}_{i}$.
In the present discussion we take the gauge-choice in analogy with the symmetric gauge of commutative gauge theory, namely, \cite{hazra}
\begin{eqnarray}
\label{e727}
\hat{A}_{i}=-\frac{\bar{B}(\theta)}{2}\epsilon_{ij}\hat{x}^j.
\end{eqnarray}
The NC magnetic field is given by
\begin{eqnarray}
\label{e164}
\hat{B}=\hat{F}_{12}=\partial_{1}\hat{A}_{2}-\partial_{2}\hat{A}_{1}-i[\hat{A}_{1},\hat{A}_{2}]_{\star}~.
\end{eqnarray}
Substituting the form of the vector potential from eq.(\ref{e727}) in eq.(\ref{e164}) yields the following expression for the NC magnetic field $\hat{B}$ in terms of $\bar{B}$
\begin{eqnarray}
\label{e280}
\hat{B}=\hat{F}_{12}=\bar{B}\left(1+\frac{\theta \bar{B}}{4}\right).
\end{eqnarray}
$\bar{B}(\theta)$ can now be fixed by comparing the above expression for the NC magnetic field with \cite{sw}
\begin{eqnarray}
\label{e290a}
\hat{B}=\frac{1}{1-\theta B}B.
\end{eqnarray}
This leads to the following expression for $\bar{B}(\theta)$ :
\begin{eqnarray}
\label{e27}
\bar{B}(\theta)=\frac{2}{\theta}[(1-\theta B)^{-1/2}-1].
\end{eqnarray}
These results will be used for the subsequent analysis of the Landau diamagnetism and the de Hass-van Alphen effect in the forthcoming sections. 


\section{Landau diamagnetism in noncommutative space}
We shall now analyze the Landau diamagnetism in NC space making use of the formalism discussed in the previous section. The analysis is carried out using coherent states \cite{feldman} which are minimal uncertainty states in the configuration space. To construct the coherent states of a charged particle in a magnetic field, we need to construct the ladder operators which will diagonalize the Hamiltonian in eq.(\ref{e101}). To do so we use the gauge choice for the vector potentials $\hat{A}_{i}$ defined in eq.(\ref{e727}).
In this gauge the Hamiltonian (\ref{e101}) takes the form
\begin{eqnarray}
\label{e2}
\hat{H}=\frac{1}{2m}\left((\hat{p}_{x}-\frac{e\bar{B}}{2c}\hat{y})^2+(\hat{p}_{y}+\frac{e\bar{B}}{2c}\hat{x})^2\right).
\end{eqnarray}
Substituting the generalized Bopp-shift transformations (\ref{e86}) and (\ref{e8zz}) in the above equation, we get an equivalent commutative Hamiltonian in terms of the commutative variables (operators) which describes the original system defined over the NC plane 
\begin{eqnarray}
\label{e700}
\hat{H}=\frac{1}{2m} \left(a^{2}p_{i}{}^2+b^{2} x_{i}{}^2 + 2 a b \epsilon_{kl}x_{k} p_{l} \right )\\
a=\frac{1}{\alpha}\left(1-\frac{e\theta\bar{B}}{4c\hbar}\right) \quad,\quad b=\frac{e\bar{B}\alpha}{2c}~. \nonumber
\end{eqnarray} 
Now we introduce the ladder operators involving the commutative phase-space variables (operators) $x$, $y$, $p_{x}$, $p_{y}$
\begin{eqnarray}
\label{e30a}
a_{x}=\frac{iap_{x}+bx}{\sqrt{2ab\hbar}}~; \quad a_{x}^{\dagger}=\frac{-iap_{x}+bx}{\sqrt{2ab\hbar}}
\end{eqnarray}
\begin{eqnarray}
\label{e31a}
a_{y}=\frac{iap_{y}+by}{\sqrt{2ab\hbar}}~; \quad  a_{y}^{\dagger}=\frac{-iap_{y}+by}{\sqrt{2ab\hbar}}
\end{eqnarray}
(where $a$ and $b$ are defined in eq.(\ref{e700})) which satisfy the commutation relations 
\begin{eqnarray}
\label{e30}
[a_{x},a_{x}^{\dagger}]=1=[a_{y},a_{y}^{\dagger}].
\end{eqnarray}
In terms of these ladder operators the Hamiltonian (\ref{e700}) can be rewritten as
\begin{eqnarray}
\label{e345}
\hat{H}=\frac{ab\hbar}{m}[(a_{x}^{\dagger} a_{x}+a_{y}^{\dagger} a_{y}+1)+i(a_{x}a_{y}^{\dagger}-a_{x}^{\dagger} a_{y})].
\end{eqnarray}
Clearly, this Hamiltonian is not in a diagonal form. To diagonalise it, we further define the pair of operators
\begin{eqnarray}
\label{e32}
a_{+}=\frac{a_{x}+ia_{y}}{\sqrt{2}}~;\quad a_{-}=\frac{a_{x}-ia_{y}}{\sqrt{2}}
\end{eqnarray}
which satisfy the commutation relations
\begin{eqnarray}
\label{e32a} 
[a_{+},a_{+}^{\dagger}]=1=[a_{-},a_{-}^{\dagger}].
\end{eqnarray}
In terms of these the Hamiltonian (\ref{e345}) can be recast in the following diagonal form
\begin{eqnarray}
\label{e33}
\hat{H}&=&\frac{2ab\hbar}{m}\left(a_{-}^{\dagger}a_{-}+\frac{1}{2}\right)\nonumber\\
&=&\left(a_{-}^{\dagger}a_{-}+\frac{1}{2}\right)\hbar\tilde{\omega}
\end{eqnarray}
where
\begin{eqnarray}
\label{e33aa}
\tilde{\omega}&=&\frac{e\bar{B}}{mc}\left(1-\frac{e\bar{B}\theta}{4c\hbar}\right)\nonumber\\
&=&\omega(\frac{2}{\theta B})\left[(1-\theta B)^{-\frac{1}{2}}-1\right]\left[1-\frac{e}{2c\hbar}\left\{(1-\theta B)^{-\frac{1}{2}}-1\right\}\right]\nonumber\\
&\equiv&\frac{e\tilde{B}}{mc}.
\end{eqnarray}
is the NC corrected frequency of the system and $\omega=\frac{eB}{mc}$ is the usual cyclotron frequency. $\tilde{B}$ is the effective magnetic field in NC space given by
\begin{eqnarray}
\label{e313a}
\tilde{B}=B(\frac{2}{\theta B})\left[(1-\theta B)^{-\frac{1}{2}}-1\right]\left[1-\frac{e}{2c\hbar}\left\{(1-\theta B)^{-\frac{1}{2}}-1\right\}\right]
\end{eqnarray}
which in the limit $\theta=0$ reduces to $\tilde{B}=B$.

\noindent Now following \cite{feldman}, we can define the following coherent state 
\begin{eqnarray}
\label{e133}
|\alpha, \xi\rangle=exp\left[-\frac{1}{4\tilde{l}^2}(|\alpha|^2+|\xi|^2)-i\sqrt{\frac{m\tilde{\omega}}{2\hbar}}a_{-}^{\dagger}+\frac{\xi \hat{X}_{-}}{2\tilde{l}^2}\right]|0\rangle
\end{eqnarray} 
where $\tilde{l}=\sqrt{\frac{\hbar}{2m\tilde{\omega}}}$ is the effective magnetic length in NC space and $\alpha$, $\xi$ are complex parameters and having dimensions of length. The operators $\hat{X}_{\pm}$ are defined as
\begin{eqnarray}
\label{e13xyz}
\hat{X}_{\pm}=\hat{X}\pm i\hat{Y}
\end{eqnarray}
where $\hat{X}$ and $\hat{Y}$ are the orbit center-coordinate operators in NC space and gets related with the phase space variables $\hat{x}_{i}$ and $\hat{p}_{i}$ through the following equations \cite{feldman}
\begin{eqnarray}
\label{e103xyz}
\hat{X}&=&\hat{x}-\frac{\hat{\pi}_{y}}{m\tilde{\omega}}\nonumber\\
\hat{Y}&=&\hat{y}+\frac{\hat{\pi}_{x}}{m\tilde{\omega}}~.
\end{eqnarray}
In the above equations $\hat{\pi}_{x}$ and $\hat{\pi}_{y}$  are the gauge invarient mechanical momenta in NC space and are defined as
\begin{eqnarray}
\label{e803xyz}
\hat{\pi}_{i}=\hat{p}_{i}+\frac{e}{c}\hat{A}_{i}~.
\end{eqnarray}
The coherent state $|\alpha,\xi\rangle$ satisfies the following eigenvalue equations
\begin{eqnarray}
\label{e13x}
a_{-}|\alpha,\xi\rangle=\sqrt{\frac{\hbar}{2m\tilde{\omega}}}\frac{\alpha}{i\tilde{l}^2}|\alpha,\xi\rangle
\end{eqnarray}
\begin{eqnarray}
\label{e131yx}
\hat{X}_{+}|\alpha,\xi\rangle=\xi|\alpha,\xi\rangle.
\end{eqnarray}
We shall now proceed to compute the
partition function of the system. The partition function is given in the standard way as
\begin{eqnarray}
\label{e183}
Z_{NC}&=&Tre^{-\beta H}\nonumber\\
&=&\sum \langle\alpha,\xi,k_{z}|e^{-\beta H}|\alpha, \xi,k_{z}\rangle\nonumber\\
&=&\sum_{k_{z}}e^{-\frac{\beta\hbar^{2}k_{z}^2}{2m}}\int\frac{d^2\xi d^2\alpha}{4\pi^2\tilde{l}^4}\langle\alpha,\xi|\left[exp\left\{-\hbar\tilde{\omega}\beta\left(a_{-}^{\dagger}a_{-}+\frac{1}{2}\right)\right\}\right]|\alpha,\xi\rangle
\end{eqnarray} 
where $\beta=\frac{1}{k_{B}T}$, $k_{B}$ is the Boltzman constant. The sum on $k_{z}$ gives the usual partion function for one-dimensional free motion along the $z$ axis.
Here we want to compute partition function for a cylindrical body of length $L$ and radius $R$, oriented along the $z$ axis. To do so we recast $Z_{NC}$ as the product of two parts : a part parallel to the magnetic field ($Z_{\parallel}$) and the other part perpendicular to the magnetic field ($Z_{\perp}$). Hence
$Z_{NC}$ can be written as 
\begin{eqnarray}
\label{e189vv}
Z_{NC}=Z_{\parallel}Z_{\perp}.
\end{eqnarray} 
By performing the sum on $k_{z}$, one can easily find that
\begin{eqnarray}
\label{e89}
Z_{\parallel}=\left(\frac{L}{h}\right)\sqrt{\frac{2\pi m}{\beta}}~.
\end{eqnarray}
Now we proceed to simplify the transverse part $Z_{\perp}$ using the coherent state approach. To do so we use the boson-operator identity
\begin{eqnarray}
\label{e89cc}
e^{xa_{-}^{\dagger}a_{-}}=\sum_{n=0}^{\infty}\frac{(e^x-1)^n}{n!}a_{-}^{\dagger n}a_{-}^n.
\end{eqnarray}
Using this identity, the term $\langle\alpha,\xi|e^{\left(-\hbar\tilde{\omega}\beta\right)a_{-}^{\dagger}a_{-}}|\alpha,\xi\rangle$ in eq.(\ref{e183}) can be simplified as
\begin{eqnarray}
\label{e80c}
\langle\alpha,\xi|e^{\left(-\hbar\tilde{\omega}\beta\right)a_{-}^{\dagger}a_{-}}|\alpha,\xi\rangle&=&\langle\alpha,\xi|1+e^{\left(-\hbar\tilde{\omega}\beta-1\right)}a_{-}^{\dagger}a_{-}+\dots|\alpha,\xi\rangle\nonumber\\
&=&1+e^{\left(-\hbar\tilde{\omega}\beta-1\right)}\langle\alpha,\xi|a_{-}^{\dagger}a_{-}|\alpha,\xi\rangle+\dots\nonumber\\
&=&1+\frac{|\alpha|^2}{2\tilde{l}^2}e^{\left(-\hbar\tilde{\omega}\beta-1\right)}+\dots\nonumber\\
&=&exp\left\{\frac{|\alpha|^2}{2\tilde{l}^2}e^{\left(-\hbar\tilde{\omega}\beta-1\right)}\right\}.
\end{eqnarray}
To simplify the above expression, we also use eq.(\ref{e13x}). Following \cite{feldman} the transverse part of the partition function $Z_{\perp}$ can now be computed using the above result and yields
\begin{eqnarray}
\label{e93c}
Z_{\perp}&=&e^{-\frac{\beta\hbar\tilde{\omega}}{2}}\int\frac{d^2\xi d^2\alpha}{4\pi^2\tilde{l}^4}exp\left\{\frac{|\alpha|^2}{2\tilde{l}^2}e^{\left(-\hbar\tilde{\omega}\beta-1\right)}\right\}\nonumber\\
&=&\left(\frac{R^2}{2\tilde{l}^2}\right)\frac{1}{e^{\frac{\beta\hbar\tilde{\omega}}{2}}-e^{-\frac{\beta\hbar\tilde{\omega}}{2}}}~.
\end{eqnarray}
Substituting eq.(\ref{e93c}) and eq.(\ref{e89}) in eq.(\ref{e189vv}), yields the following expression for the partition function in NC space
\begin{eqnarray}
\label{e189}
Z_{NC}=\frac{V}{h}\left(\frac{2\pi m}{\beta}\right)^{\frac{1}{2}}\left(\frac{m\tilde{\omega}}{4\pi\hbar}\right)\frac{1}{\sinh\left(\frac{\beta\hbar\tilde{\omega}}{2}\right)}
\end{eqnarray}  
where $V=\pi R^{2}L$ is the volume of the cylinder. From the above result, we can therefore construct the free energy of the system to be
\begin{eqnarray}
\label{e109}
F_{NC}=-\frac{n}{\beta}ln Z_{NC}
\end{eqnarray}
where $n$ is the number density of electrons. Hence the magnetization of the system is given by
\begin{eqnarray}
\label{e109x}
M_{NC}&=&-\frac{\partial F_{NC}}{\partial B}\nonumber\\
&=&\frac{ne\hbar}{2mc}\left(1-\theta B\right)^{-\frac{3}{2}}\left[1-\frac{e}{c\hbar}\left\{(1-\theta B)^{-\frac{1}{2}}-1\right\}\right]\left[\frac{2}{\beta\hbar\tilde{\omega}}-coth\left(\frac{\beta\hbar\tilde{\omega}}{2}\right)\right].
\end{eqnarray}
In the limit $\theta=0$, we have $\tilde{\omega}=\omega$ since the factor $\left[1-\frac{e}{c\hbar}\left\{(1-\theta B)^{-\frac{1}{2}}-1\right\}\right]=1$. Therefore we recover the commutative result for the magnetization of the electron gas in the presence of a magnetic field \cite{feldman}
\begin{eqnarray}
\label{e409}
M=\frac{ne\hbar}{2mc}\left[\frac{2}{\beta\hbar\omega}-coth\left(\frac{\beta\hbar\omega}{2}\right)\right].
\end{eqnarray}
The magnetization in eq.(\ref{e109x}) can be expressed in a more convenient way as
\begin{eqnarray}
\label{e412}
M_{NC}=-\frac{ne\hbar}{2mc}\left(1-\theta B\right)^{-\frac{3}{2}}\left[1-\frac{e}{c\hbar}\left\{(1-\theta B)^{-\frac{1}{2}}-1\right\}\right]L\left(\frac{\beta\hbar\tilde{\omega}}{2}\right)
\end{eqnarray}
where $L\left(\frac{\beta\hbar\tilde{\omega}}{2}\right)=\left[\frac{2}{\beta\hbar\tilde{\omega}}-coth\left(\frac{\beta\hbar\tilde{\omega}}{2}\right)\right]$ is the Langevin function of $\frac{\beta\hbar\tilde{\omega}}{2}$. In the high temperature regime, $\beta<<1$, and hence we have $L\left(\frac{\beta\hbar\tilde{\omega}}{2}\right)\simeq\frac{1}{3}\frac{\beta\hbar\tilde{\omega}}{2}$. The magnetization in eq.(\ref{e412}) therefore reduces to the following form
\begin{eqnarray}
\label{e124}
M_{NC}=-\frac{1}{3}\left(\frac{\beta\hbar\tilde{\omega}}{2}\right)\frac{ne\hbar}{2mc}\left(1-\theta B\right)^{-\frac{3}{2}}\left[1-\frac{e}{c\hbar}\left\{(1-\theta B)^{-\frac{1}{2}}-1\right\}\right].
\end{eqnarray} 
We are now in a position to compute the susceptibility of the Landau system in NC space. With the usual definition, the magnetic susceptibility of the Landau system in NC space is given by 
\begin{eqnarray}
\label{e124b}
\chi_{NC}&=&\frac{1}{n}\frac{\partial M_{NC}}{\partial B}\nonumber\\
&=&\chi_{L}(1-\theta B)^{-3}\left[(1-\theta B)^{\frac{1}{2}}\left(1-\frac{e}{ch}\left\{(1-\theta B)^{-\frac{1}{2}}-1\right\}\right)\right.\nonumber\\
&&\left.\times\left((1-\theta B)^{-\frac{1}{2}}-1\right)\left(1-\frac{e}{2ch}\left\{(1-\theta B)^{-\frac{1}{2}}-1\right\}\right)-\frac{e}{ch}\left\{(1-\theta B)^{-\frac{1}{2}}-1\right\}\right.\nonumber\\
&&\left.\times\left(1-\frac{e}{2ch}\left\{(1-\theta B)^{-\frac{1}{2}}-1\right\}\right)\right.\nonumber\\
&&\left.+\left(1-\frac{e}{ch}\left\{(1-\theta B)^{-\frac{1}{2}}-1\right\}\right)\left(1-\frac{e}{2ch}\left\{(1-\theta B)^{-\frac{1}{2}}-1\right\}\right)-\frac{e}{2ch}\left\{(1-\theta B)^{-\frac{1}{2}}-1\right\}\right.\nonumber\\
&&\left.\left(1-\frac{e}{ch}\left\{(1-\theta B)^{-\frac{1}{2}}-1\right\}\right)\right]
\end{eqnarray} 
where $\chi_{L}=-\frac{1}{3}\left(\frac{e\hbar}{2mc}\right)^2\beta$ is the usual Landau diamagnetic susceptibility in commutative space. From the above expression we observe that the susceptibility picks up the NC correction. Our result also captures the effect of noncommutativity upto all orders in $\theta$ which is in contrast with the result obtained in \cite{jellal} where the NC correction involves only up to second-order in $\theta$. However, the result is independent of the scaling parameter $\alpha$ appearing in the expression of the generalized Bopp-shift transformations (\ref{e86}) and (\ref{e8zz}). Another important result that we get from the above analysis is that the Landau susceptibility in NC space depends on the magnetic field. This is in contrast to the commutative result where the Landau susceptibility is independent of the magnetic field. It is reassuring to note that in the limit $\theta=0$, the commutative result is recovered.

\section{de Hass--van Alphen effect in noncommutative space}
The de Haas-van Alphen (dHvA) effect is an oscillatory variation of the diamagnetic susceptibility as a function of the magnetic field strength ($B$). 
The effect was first observed by de Haas and van Alphen in (1930). 
They measured the
magnetization $M$ of semimetal bismuth (Bi) as a function of the magnetic field ($B$) at 14.2 K and found that the magnetic susceptibility $M/B$ is a periodic function
of the reciprocal of the magnetic field ($1/B$). This phenomenon is observed only at low temperatures and high magnetic fields. It was Landau who explained this phenomena as a direct consequence of the quantization of the motion of the electron in a magnetic field.

Here we shall study the dHvA effect in NC space. To do so we shall first calculate the total energy $E$ of $N$ electrons moving in a NC plane in the presence of a magnetic field perpendicular to the plane. The energy of a single electron in a state $j$ is given by eq.(s)(\ref{e33}) and (\ref{e33aa}) 
\begin{eqnarray}
\label{e4y}
E_{j}=\left(j+\frac{1}{2}\right)2\mu_{B}\tilde{B}~;~j=0,1,2,...~.
\end{eqnarray} 
The degeneracy factor (including the spin degeneracy) of the Landau system in NC space is given by
\begin{eqnarray}
\label{e4yy}
g&=&2\frac{eA\tilde{B}}{hc}\nonumber\\
&=&\frac{N\tilde{B}}{B_{0}}
\end{eqnarray} 
where $A$ is the area of the two dimensional plane, $B_{0}=\frac{nhc}{2e}$ with $n=\frac{N}{A}$ being the number density of the electrons.
At $T=0$, all the electrons tend to settle in the lowest Landau level. Now from eq.(\ref{e4yy}), we see that if $\tilde{B}>B_{0}$, then $g>N$ and hence all the electrons can be accommodated in the lowest Landau level. Therefore in this case the total energy of the $N$ electrons is given by
\begin{eqnarray}
\label{e49y}
E_{tot}^{(0)}=N\mu_{B}\tilde{B}.
\end{eqnarray} 
This is the ground state energy of all the $N$ electrons. However for $\tilde{B}<B_{0}$, not all the electrons can be accommodated in the ground state. Suppose that the magnetic field is such that the $j$-th Landau level is completely filled while some are in the next level, that is the $(j+1)$-th level. In this case we can write
\begin{eqnarray}
\label{e49xx}
(j+1)g<N<(j+2)g
\end{eqnarray} 
which in turn leads to
\begin{eqnarray}
\label{e49kxx}
\frac{1}{j+2}<\frac{\tilde{B}}{B_{0}}<\frac{1}{j+1}~.
\end{eqnarray} 
In this case the total energy of the system of $N$ electrons is given by
\begin{eqnarray}
\label{e34xx}
E_{tot}^{(j)}&=&g\sum_{i=0}^{j}\left[(2i+1)\mu_{B}\tilde{B}+\left\{N-g(i+1)\right\}(2i+3)\mu_{B}\tilde{B}\right]\nonumber\\
&=&N\mu_{B}B_{0}\left[(2j+3)\frac{\tilde{B}}{B_{0}}-(j+1)(j+2)\left(\frac{\tilde{B}}{B_{0}}\right)^2\right].
\end{eqnarray} 
The magnetization of the system can be calculated from the eq.(s)(\ref{e49y}) and (\ref{e34xx}) by the usual relation
$M_{NC}=-\frac{\partial E}{\partial B}$ and yields
\begin{eqnarray}
\label{e34x}
M_{NC}=-N\mu_{B}(1-\theta B)^{-\frac{3}{2}}\left[1-\frac{e}{ch}\left\{(1-\theta B)^{-\frac{1}{2}}-1\right\}\right]~;~\tilde{B}>B_{0}
\end{eqnarray} 
\begin{eqnarray}
\label{e3001}
M_{NC}=-N\mu_{B}(1-\theta B)^{-\frac{3}{2}}\left[1-\frac{e}{ch}\left\{(1-\theta B)^{-\frac{1}{2}}-1\right\}\right]\left[(2j+3)-2(j+1)(j+2)\frac{\tilde{B}}{B_{0}}\right]~;\nonumber\\ 
\frac{1}{j+2}<\frac{\tilde{B}}{B_{0}}<\frac{1}{j+1}~.
\end{eqnarray}
Hence the susceptibilities of the Landau system for the two different range of the magnetic field are given by
\begin{eqnarray}
\label{e34h}
\chi_{NC}=-\left(\frac{N\mu_{B}\theta}{2}\right)(1-\theta B)^{-3}\left[3(1-\theta B)^{\frac{1}{2}}\left(1-\frac{e}{ch}\left\{(1-\theta B)^{-\frac{1}{2}}-1\right\}\right)-\frac{e}{c\hbar}\right]~;\nonumber\\
\tilde{B}>B_{0}
\end{eqnarray}  
\begin{eqnarray}
\label{e301x}
\chi_{NC}&=&-N\mu_{B}(1-\theta B)^{-3}\left[\frac{\theta}{2}\left(3(1-\theta B)^{\frac{1}{2}}\left\{1-\frac{e}{ch}\left((1-\theta B)^{-\frac{1}{2}}-1\right)\right\}-\frac{e}{c\hbar}\right)\right.\nonumber\\
&&\left.\times\left\{(2j+3)-2(j+1)(j+2)\frac{\tilde{B}}{B_{0}}\right\}-\left(\frac{2}{B_{0}}\right)(j+1)(j+2)\right.\nonumber\\
&&\left.\times\left\{1-\frac{e}{ch}\left\{(1-\theta B)^{-\frac{1}{2}}-1\right\}\right\}^2\right]~;
~\frac{1}{j+2}<\frac{\tilde{B}}{B_{0}}<\frac{1}{j+1}~.
\end{eqnarray}
The above analysis leads us to conclude that in NC space the oscillation of the magnetization and that of the magnetic susceptibility gets NC corrections. Our results are valid upto all orders in $\theta$. Interestingly we have found that the susceptibility is nonzero in the range $\tilde{B}>B_{0}$ which is in contrast with the result in commutative space where $\chi=0$ for $B>B_{0}$. This is a new result in this paper. Once again the commutative results can be recovered in the $\theta=0$ limit.

\section{Conclusions}
In this paper we have investigated the effect of  noncommutativity on the thermodynamics of a charged particle moving in a two dimensional plane in the presence of a background magnetic field perpendicular to the plane.  We have adopted an approach where we map the NC problem to an equivalent commutative problem by using a generalized Bopp-shift transformation. We have also incorporated the Seiberg-Witten map to correctly consider the issues of gauge invariance involved in noncommutative theories. The commutative equivalent Hamiltonian in terms of commutative variables and NC parameter $\theta$ is then obtained. With this formalism in hand, 
we compute the partition function of the Landau system using the coherent state approach and from this we calculate the magnetization and the magnetic susceptibility in the high temperature limit. It is observed that the results get modified by the noncommutative parameter $\theta$. Our results are valid upto all orders in the noncommutative parameter $\theta$ which is in contrast with those in the existing literature where the result is valid only up to second order in $\theta$. We then study the de Hass-van Alphen effect in NC space in the low temperature limit and find that the oscillation of the magnetic susceptibility with respect to the applied magnetic field gets corrected due to noncommutativity.  Interestingly, we find that the susceptibility in the noncommutative scenario is non-zero in the range of the magnetic field greater than the threshold value. This is in complete contrast to its commutative counterpart where the susceptibility vanishes for this range of the magnetic field. As a future work, our aim is to extend our analysis in noncommutative phase-space.

 
\section*{Acknowledgement}
S.G. acknowledges the support by DST SERB under Start Up Research Grant (Young Scientist), File No.YSS/2014/000180.


\end{document}